\documentclass[letterpaper,10pt]{article}
\usepackage{amsmath,mathptm,graphicx,eprint}

\setreportheading{Masahiko Yoshioka}{PREPRINT}

\begin{document}

\title{Chaos synchronization in gap-junction-coupled neurons}
{\date{{June 10, 2004}\\ {\normalsize ({Revised on May 9, 2005})}}}
\author{{{Masahiko Yoshioka}}\thanks{Electronic address: {{myosioka@brain.riken.go.jp}}}\\
\begin{minipage}{1.0\textwidth}{\normalsize\it\begin{center}\ \\ {{%
Brain Science Institute, The Institute of Physical and Chemical Research (RIKEN)\\
Hirosawa 2-1, Wako-shi, Saitama, 351-0198, Japan}}\end{center}}\end{minipage}}

\maketitle

\begin{abstract}

Depending on temperature the modified Hodgkin-Huxley (MHH) equations
exhibit a variety of dynamical behavior including intrinsic chaotic
firing. We analyze synchronization in a large ensemble of MHH neurons
that are interconnected with gap junctions. By evaluating tangential
Lyapunov exponents we clarify whether synchronous state of neurons is
chaotic or periodic. Then, we evaluate transversal Lyapunov exponents
to elucidate if this synchronous state is stable against infinitesimal
perturbations. Our analysis elucidates that with weak gap junctions,
stability of synchronization of MHH neurons shows rather complicated
change with temperature.  We, however, find that with strong gap
junctions, synchronous state is stable over the wide range of
temperature irrespective of whether synchronous state is chaotic or
periodic. It turns out that strong gap junctions realize the robust
synchronization mechanism, which well explains synchronization in
interneurons in the real nervous system.

\end{abstract}

{

\vspace{0.2em}

\noindent {\footnotesize PACS numbers: {87.18.Sn,02.30.Oz,05.45.Xt,05.45.Ra}}

\vspace{1em}

}

\noindent In the rat hippocampus interneurons show the high frequency synchronization
during the gamma oscillation ($\sim$40Hz) and sharp wave burst
($\sim$200Hz)\cite{buzsaki3}, and such simple synchronization of a large
ensemble of neurons has attracted much attention of theoretical
researchers\cite{wang,golomb,whittington1,wang2,chow,tiesinga,myosioka7,ermentrout,hansel}.
One major analysis for these studies is the phase reduction
method\cite{kuramoto,ermentrout,hansel}, in which phase variables are
utilized to represent the periodic behavior of neurons.  The phase
reduction method is, however, applicable only to the case of
infinitesimal interactions.  Moreover, if neurons behave aperiodic, phase variables are indefinable.  The general synchronization properties of
strongly coupled neurons thus remained unclear, especially in the case
of chaotic neurons.

Meanwhile, studies of synchronization of a large ensemble of chaotic
oscillators have made a remarkable progress in recent
years\cite{fujisaka,kaneko,maistrenko,pikovsky}.  The major targets of
these studies are simple chaotic oscillators such as Lorenz equations
and logistic maps.  Synchronous state of these oscillators is
characterized by two types of Lyapunov exponents: tangential Lyapunov
exponents and transversal Lyapunov exponents.  While tangential Lyapunov
exponents clarify whether synchronous state is chaotic or periodic,
transversal Lyapunov exponents elucidate if synchronous state is stable
against infinitesimal perturbations.  In the present study, we employ
these sophisticated techniques in chaos synchronization theory
to investigate synchronization of neurons.  We
show that tangential and transversal Lyapunov exponents enable us to analyze stability of
synchronization in a large ensemble of neurons for arbitrary neuron
dynamics and arbitrary strength of interactions.

\begin{figure}[h]
 {}{{}}{}{}
\begin{center}
\includegraphics[scale=1.0]{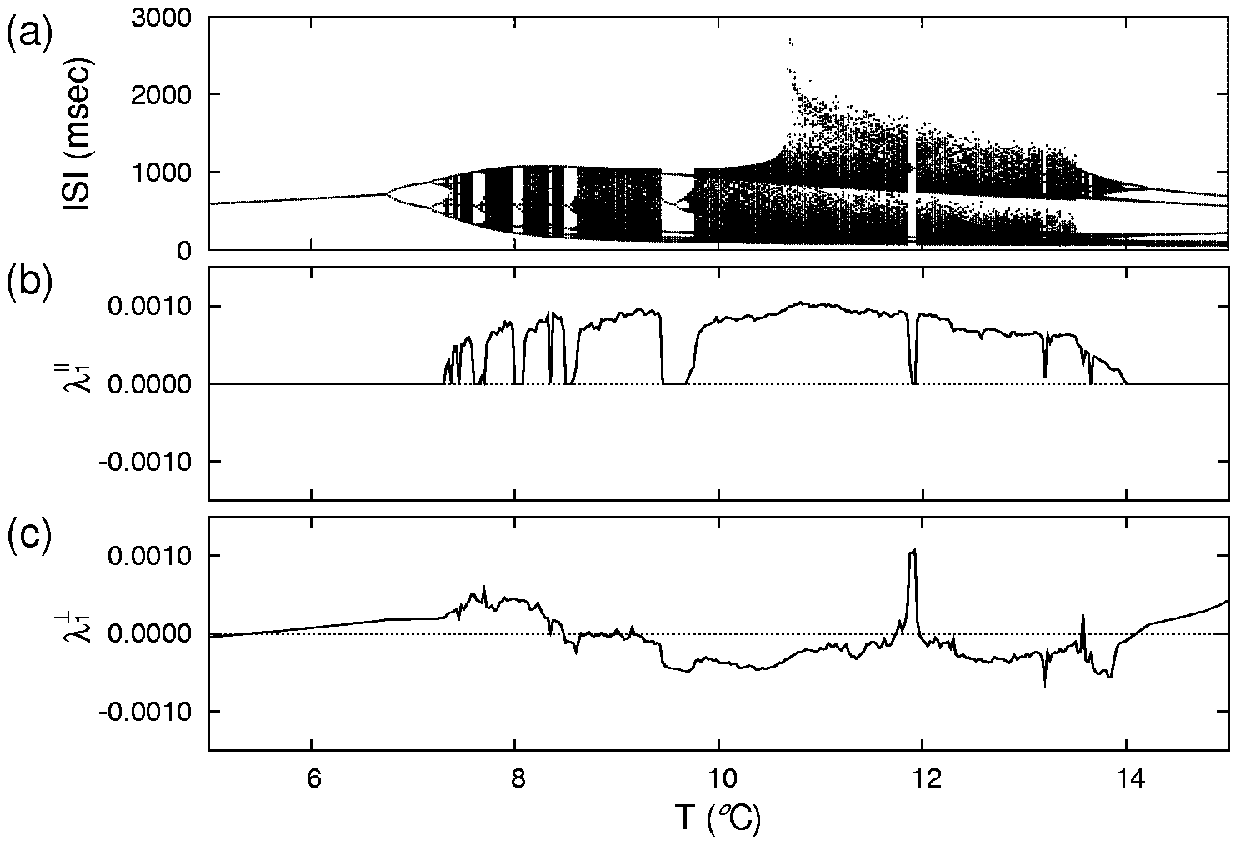}
\end{center}
 \caption{{(a)}\ The interspike intervals (ISIs) in stationary state of
 an isolated single MHH neuron are plotted as a function of
 temperature~$T$.  The ISI distribution of synchronous state of $N$
 neurons ${{\bf x}}^\ast$ is the same as a isolated single neuron since the
 interaction term in Eq.~(\ref{eq:sync}) vanishes in the present system.
 {(b)}\ The largest tangential Lyapunov exponent~${\lambda^{\mbox{\tiny $\parallel$}}_{1}}$, which
 characterizes synchronous state~${{\bf x}}^\ast$ described in {(a)}, is
 plotted. In the present system, ${\lambda^{\mbox{\tiny $\parallel$}}_{1}}$ is independent of $g$.
 {(c)}\ The largest transversal Lyapunov exponent~${\lambda^{\mbox{\tiny $\perp$}}_{1}}$ is plotted
 for $g=0.02$~{$\mbox{mS/cm}^2$}. When ${\lambda^{\mbox{\tiny $\perp$}}_{1}}$ takes a negative value,
 synchronization is stable.  The sampling points are $T=5+0.025\times
 k\mbox{({\char'27\kern-.3em\hbox{C}})}\ (k=0,1,2,\ldots)$. }\label{fig:result}
\end{figure}

The concrete target of the present analysis is a network of $N(\ge 2)$
spiking neurons that obey the modified Hodgkin-Huxley (MHH)
equations\cite{braun,feudel}.  The MHH equations are four-dimensional
nonlinear differential equations, which include temperature-dependent
scaling factors $\rho=A_1^{(T-T_0)/10}$ and $\phi=A_2^{(T-T_0)/10}$ (See Ref.~\cite{feudel}.)
With $T$ changing, a MHH neuron shows a variety of
dynamical behavior including chaotic firing as shown in
Fig.~\ref{fig:result}{(a)} (ISI and so on will be explained later.)
For the sake of simplicity, we denote this MHH neuron dynamics by
$d{{\bf x}}/dt={{\bf F}}({{\bf x}})$ with neuron state vector
${{\bf x}}={\left ( {v,w_{1},\ldots,w_{n-1}} \right )}^{\rm T}$, where $v$ represents the
membrane potential and ${\left\{ {w_l} \right \}}$ describe gating of ion channels.
We assume that $N$~neurons ${\left\{ {{{\bf x}}_i} \right \}}$ are interconnected with
all-to-all gap junctions.  Since gap junctions induce electric currents
proportional to potential difference between neurons, the dynamics of
the neural networks is expressed as
\begin{equation}
\frac{d{{{\bf x}}}_i}{dt}={{\bf F}}({{\bf x}}_i)+(I_i/c,0,\ldots,0)^{\rm T},\ i=1,\ldots,N \label{eq:mhh}
\end{equation}
with
\begin{equation}
 I_i=\frac{g}{N}\sum_j{\left ( {v_j-v_i} \right )}=\frac{g}{N}\sum_j{\left ( {x_{j1}-x_{i1}} \right )},\label{eq:interaction}
\end{equation}
where constant $c=1.0\ \mu {\rm F}/{\rm cm}^2$ is capacitance of the
membrane and $I_i$ is the electric current induced by gap junctions.

The above dynamics can be generalized to the form
\begin{equation}
\frac{d{{{\bf x}}}_i}{dt}={{\bf F}}({{\bf x}}_i)+\frac{g}{N}\sum_j{{\bf G}}({{\bf x}}_i,{{\bf x}}_j).\label{eq:general}
\end{equation}
Therefore, we investigate synchronous state in this general mean-field
dynamics.  We assume stationary synchronous state
${{\bf x}}_1^\ast=\ldots={{\bf x}}_N^\ast={{\bf x}}^\ast$, which obeys
\begin{equation}
\frac{d {{{\bf x}}}^\ast}{dt}={{\bf F}}({{\bf x}}^\ast)+g{{\bf G}}({{\bf x}}^\ast,{{\bf x}}^\ast).\label{eq:sync}
\end{equation}
To elucidate the stability of this synchronous state we investigate
perturbed state ${{\bf x}}_i={{\bf x}}^\ast+\delta{{\bf x}}_i$.
We define Jacobi matrices such that
${{\bf F}}({{\bf x}}^\ast+\delta{{\bf x}})={{\bf F}}({{\bf x}}^\ast)+{{\bf F}}^\prime({{\bf x}}^\ast)\delta{{\bf x}}+(\mbox{higher
order})$ and
${{\bf G}}({{\bf x}}^\ast+\delta{{\bf x}}_1,{{\bf x}}^\ast+\delta{{\bf x}}_2)={{\bf G}}({{\bf x}}^\ast,{{\bf x}}^\ast)+{{\bf G}}^\prime_1({{\bf x}}^\ast,{{\bf x}}^\ast)\delta{{\bf x}}_1+{{\bf G}}^\prime_2({{\bf x}}^\ast,{{\bf x}}^\ast)\delta{{\bf x}}_2+(\mbox{higher
order})$.
Then, Taylor series expansion to the first order yields
\begin{eqnarray}
 \frac{d(\delta{{{\bf x}}}_i)}{dt}&=&{\left [ {{{\bf F}}^\prime({{\bf x}}^\ast)+g{{\bf G}}^\prime_1({{\bf x}}^\ast,{{\bf x}}^\ast)} \right ]}\delta{{\bf x}}_i\nonumber\\
 &&+g{{\bf G}}^\prime_2({{\bf x}}^\ast,{{\bf x}}^\ast)\frac{1}{N}\sum_j\delta{{\bf x}}_j.\label{eq:perturbation}
\end{eqnarray}
The naive evaluation of this $N$-body dynamics brings about an eigenvalue
problem with the large size of matrix.
We hence define mean state
$\overline{{{\bf x}}}=(1/N)\sum_i{{\bf x}}_i$ and obtain the dynamics of its deviation~$\delta\overline{{{\bf x}}}=(1/N)\sum_i\delta{{\bf x}}_i$ in the closed form
\begin{equation}
 \frac{d(\delta{\overline{{{\bf x}}}})}{dt}={\left [ {{{\bf F}}^\prime({{\bf x}}^\ast)+g{{\bf G}}^\prime_1({{\bf x}}^\ast,{{\bf x}}^\ast)+g{{\bf G}}^\prime_2({{\bf x}}^\ast,{{\bf x}}^\ast)} \right ]}\delta\overline{{{\bf x}}}.\label{eq:dave}
\end{equation}
For this $n$-dimensional linear dynamics, we can
define the spectrum of $n$~Lyapunov exponents~${\{ {\lambda^{\mbox{\tiny $\parallel$}}_{l}}\}}_{l=1,\ldots,n}$. These exponents are the so-called
tangential Lyapunov exponents.
To the first order, Eqs.~(\ref{eq:sync}) and (\ref{eq:dave}) are equivalent to
\begin{equation}
\frac{d}{dt}{({{\bf x}}^\ast+\delta\overline{{{\bf x}}})}={{\bf F}}({{\bf x}}^\ast+\delta\overline{{{\bf x}}})+g{{\bf G}}({{\bf x}}^\ast+\delta\overline{{{\bf x}}},{{\bf x}}^\ast+\delta\overline{{{\bf x}}}).\label{eq:ave}
\end{equation}
Solving Eqs.~(\ref{eq:sync}) and (\ref{eq:ave}) numerically we can
calculate time evolution of sufficiently small
deviation~$\delta\overline{{{\bf x}}}$.  Evaluating this time evolution of $\delta\overline{{{\bf x}}}$ by the
well-known computational method for Lyapunov exponents\cite{shimada} we
can calculate ${\{ {\lambda^{\mbox{\tiny $\parallel$}}_{l}}\}}$ numerically.
Note that in this calculation of ${\{ {\lambda^{\mbox{\tiny $\parallel$}}_{l}}\}}$ we do not have to solve the huge $N$-body dynamics in Eq.~(\ref{eq:general}).
Since replacement of ${{\bf x}}^\ast$ in
Eq.~(\ref{eq:sync}) by ${{\bf x}}^\ast+\delta\overline{{{\bf x}}}$ gives the same
dynamics as Eq.~(\ref{eq:ave}), ${\{ {\lambda^{\mbox{\tiny $\parallel$}}_{l}}\}}$ indicate the characteristics
of synchronous state~${{\bf x}}^\ast$, that is, when
synchronous state~${{\bf x}}^\ast$ is periodic (chaotic), the largest tangential Lyapunov exponent~${\lambda^{\mbox{\tiny $\parallel$}}_{1}}$ takes the zero value (a positive value).

We have evaluated mean state~$\overline{{{\bf x}}}$ by tangential Lyapunov
exponents~${\{ {\lambda^{\mbox{\tiny $\parallel$}}_{l}}\}}$. We now
investigate 
deviations around mean state: ${{\bf x}}_i=\overline{{{\bf x}}}+\delta\tilde{{{\bf x}}}_i$.
Subtracting Eq.~(\ref{eq:dave}) from Eq.~(\ref{eq:perturbation}) we
obtain the dynamics of $\delta\tilde{{{\bf x}}}_i$ in the closed form
\begin{equation}
 \frac{d(\delta{\tilde{{{\bf x}}}}_i)}{dt}={\left [ {{{\bf F}}^\prime({{\bf x}}^\ast)+g{{\bf G}}^\prime_1({{\bf x}}^\ast,{{\bf x}}^\ast)} \right ]}\delta\tilde{{{\bf x}}}_i.\label{eq:dtilde}
\end{equation}
When synchronization is stable, all deviations
${\left\{ {\delta\tilde{{{\bf x}}}_i} \right \}}$ must converge to ${\left\{ {{\bf 0}} \right \}}$.
Since these $N$ dynamics of deviations are completely identical, it
suffices to evaluate only one dynamics among them.  For $n$-dimensional
linear dynamics in Eq.~(\ref{eq:dtilde}) we can define the spectrum of
$n$~Lyapunov exponents~${\{ {\lambda^{\mbox{\tiny $\perp$}}_{l}}\}}_{l=1,\ldots,n}$. These exponents are the
so-called transversal Lyapunov exponents.  The largest transversal
Lyapunov exponent~${\lambda^{\mbox{\tiny $\perp$}}_{1}}$ takes a negative value when synchronous
state is stable in the sense of Milnor\cite{maistrenko,milnor}.  To the
first order, Eqs.~(\ref{eq:sync}) and (\ref{eq:dtilde}) are equivalent
to
\begin{equation}
\frac{d}{dt}{({{\bf x}}^\ast+\delta\tilde{{{\bf x}}}_i)}={{\bf F}}({{\bf x}}^\ast+\delta\tilde{{{\bf x}}}_i)+g{{\bf G}}({{\bf x}}^\ast+\delta\tilde{{{\bf x}}}_i,{{\bf x}}^\ast).\label{eq:tilde}
\end{equation}
Applying the computational method for Lyapunov exponents\cite{shimada} to 
Eqs.~(\ref{eq:sync}) and (\ref{eq:tilde}), we can calculate ${\{ {\lambda^{\mbox{\tiny $\perp$}}_{l}}\}}$ numerically.
 
Let us apply the above analysis to investigating synchronization in
networks of MHH neurons defined by Eqs.~(\ref{eq:mhh}) and (\ref{eq:interaction}). First, we 
calculate synchronous state ${{\bf x}}^\ast$ in Eq.~(\ref{eq:sync}).
Note that in the present system the interaction term
$g{{\bf G}}({{\bf x}}^\ast,{{\bf x}}^\ast)$ in Eq.~(\ref{eq:sync}) vanishes because of
Eq.~(\ref{eq:interaction}).
Therefore, the behavior of ${{\bf x}}^\ast$ in Eq.~(\ref{eq:sync}) is
completely the same as that of an isolated single MHH neuron.
For the
rough illustration of a single MHH neuron behavior, we define
the $k$-th spike timing $t(k)$ by the time when membrane potential
$v^\ast=x_1^\ast$ crosses the threshold value $\theta=-20$~mV from
below, and then calculate interspike intervals (ISIs) $t(k+1)-t(k)\
(k=1,2,\ldots)$ in Fig.~\ref{fig:result}{(a)}.  Below $T=6.8${\char'27\kern-.3em\hbox{C}}, ISIs
take the single value around 650~msec, implying the simple periodic
firing in which only one spike arises during the period.  At $T=6.8${\char'27\kern-.3em\hbox{C}},
however, period doubling bifurcation occurs so that the neuron fires
twice during the period.  After that, following typical period doubling
cascade, the MHH neuron dynamics reaches the chaotic regime beyond
$T=7.3${\char'27\kern-.3em\hbox{C}}, where ISI distribution becomes blurred.  In this chaotic
regime, we, however, observe several periodic windows, in which the
behavior of neuron becomes periodic abruptly.

Second, from Eqs.~(\ref{eq:sync}) and (\ref{eq:ave}), we calculate the
tangential Lyapunov exponents~${\{ {\lambda^{\mbox{\tiny $\parallel$}}_{l}}\}}$ for the exact characterization of
synchronous state~${{\bf x}}^\ast$ illustrated in
Fig.~\ref{fig:result}{(a)}. In the present system, ${\{ {\lambda^{\mbox{\tiny $\parallel$}}_{l}}\}}$ are
independent of $g$ because of the interaction in
Eq.~(\ref{eq:interaction}).  In Fig.~\ref{fig:result}{(b)}, we plot the
largest tangential Lyapunov exponents~${\lambda^{\mbox{\tiny $\parallel$}}_{1}}$ as a function of
temperature $T$.  When synchronous state~${{\bf x}}^\ast$ in
Fig.~\ref{fig:result}{(a)}\ is periodic, ${\lambda^{\mbox{\tiny $\parallel$}}_{1}}$ takes the zero
value.  In chaotic regime beyond $T=7.3{\char'27\kern-.3em\hbox{C}}$, ${\lambda^{\mbox{\tiny $\parallel$}}_{1}}$ takes a positive
value, though we observe the several valleys of ${\lambda^{\mbox{\tiny $\parallel$}}_{1}}$ corresponding
to the periodic windows observed in Fig.~\ref{fig:result}{(a)}.

Third, we calculate the transversal Lyapunov exponents~${\{ {\lambda^{\mbox{\tiny $\perp$}}_{l}}\}}$
 from Eqs.~(\ref{eq:sync}) and (\ref{eq:tilde}).  
${\{ {\lambda^{\mbox{\tiny $\perp$}}_{l}}\}}$ depend on parameter~$g$.
In Fig.~\ref{fig:result}{(c)}\ 
we calculate the largest transversal Lyapunov exponent~${\lambda^{\mbox{\tiny $\perp$}}_{1}}$ as a
function of temperature~$T$ for $g=0.02$~{$\mbox{mS/cm}^2$}.  When ${\lambda^{\mbox{\tiny $\perp$}}_{1}}$ takes a
negative value, synchronization of MHH neurons can occur.

\begin{figure}[h]
 {}{{}}{}{}
\begin{center}
\includegraphics[scale=1.4]{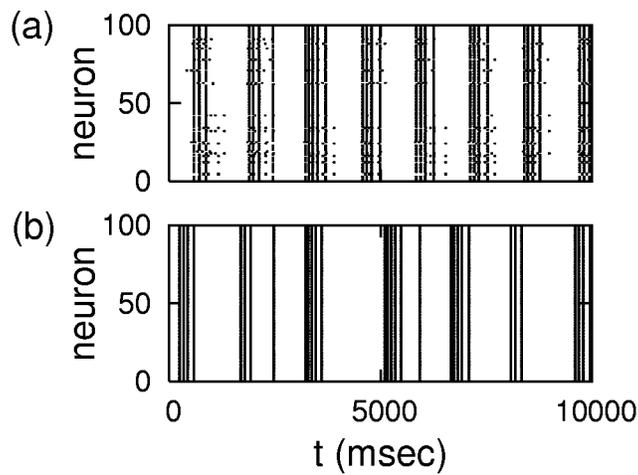}
\end{center}
 \caption{The results of numerical simulations of 100 MHH neurons
 with $g=0.02$~{$\mbox{mS/cm}^2$} are plotted for {(a)}\
 $T=11.9${\char'27\kern-.3em\hbox{C}}\ and {(b)}\ $T=12.1${\char'27\kern-.3em\hbox{C}}.
 Each dot represents spike timing in stationary state realized after $t=1.0\times 10^7$~msec.}\label{fig:simulation}
\end{figure}

When we assume weak gap junctions as in Fig.~\ref{fig:result}($g=0.02$~{$\mbox{mS/cm}^2$}),
the condition for synchronization of MHH neurons is rather complicated.
Periodic synchronous state is stable in some conditions and unstable in
other conditions. We also see stable chaotic synchronous state in
some values of temperature~$T$.
Around $T\sim 12${\char'27\kern-.3em\hbox{C}}\ we find unstable periodic synchronous state inside
the periodic window (${\lambda^{\mbox{\tiny $\parallel$}}_{1}}=0$ and $0<{\lambda^{\mbox{\tiny $\perp$}}_{1}}$ at $T=11.9${\char'27\kern-.3em\hbox{C}}) while we find stable chaotic synchronous state
outside the periodic window ($0<{\lambda^{\mbox{\tiny $\parallel$}}_{1}}$ and ${\lambda^{\mbox{\tiny $\perp$}}_{1}}<0$ at $T=12.1${\char'27\kern-.3em\hbox{C}}). Actually, 
the numerical simulations of 100 MHH neurons in Fig~\ref{fig:simulation}
show the good agreement with the results of our analysis.
Around $T\sim 9.5${\char'27\kern-.3em\hbox{C}}, however, synchronous state is stable both inside
and outside the periodic window.  With weak gap junctions the condition
for synchronization is so complicated that its intuitive explanation
is difficult.

\begin{figure}[h]
 {}{{}}{}{}
\begin{center}
\includegraphics[scale=1.4]{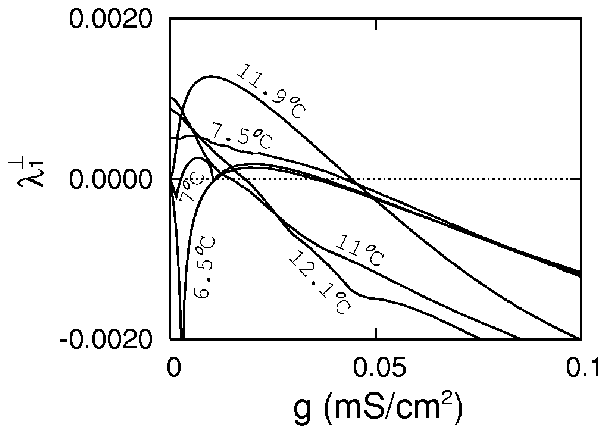}
\end{center}
 \caption{The largest transversal Lyapunov exponent~${\lambda^{\mbox{\tiny $\perp$}}_{1}}$ is plotted as a
 function of $g$ for various values of temperature~$T$. The numbers in the
 figure indicate temperature~$T$.}\label{fig:largeg}
\end{figure}

On the other hand, with strong gap junctions, synchronization of MHH
neurons is stable over the wide range of temperature~$T$.  In
Fig.~\ref{fig:largeg}, we plot the largest transversal Lyapunov
exponent~${\lambda^{\mbox{\tiny $\perp$}}_{1}}$ as a function of $g$ for various values of
temperature~$T$.  Eqs.~(\ref{eq:dave}) and (\ref{eq:dtilde}) show that
when parameter~$g$ takes the zero value, transversal Lyapunov
exponents~${\{ {\lambda^{\mbox{\tiny $\parallel$}}_{l}}\}}$ take the same value as tangential Lyapunov
exponents~${\{ {\lambda^{\mbox{\tiny $\perp$}}_{l}}\}}$.  Therefore, synchronous state in
Fig.~\ref{fig:largeg} is periodic for $T=6.5,\ 7,\mbox{ and }11.9$~({\char'27\kern-.3em\hbox{C}})
and chaotic for $T=7.5,\ 11,\mbox{ and }12.1$~({\char'27\kern-.3em\hbox{C}}).  In these six
temperature, the behavior of MHH neurons are quite different from one
another.  However, all ${\lambda^{\mbox{\tiny $\perp$}}_{1}}$ take negative values if we
increase the strength of gap junctions beyond $g=0.05$.
In all the temperature we investigate ($5{\char'27\kern-.3em\hbox{C}}<T<15{\char'27\kern-.3em\hbox{C}}$), we find a
certain value of $g$ beyond which ${\lambda^{\mbox{\tiny $\perp$}}_{1}}$ always take a negative
value.  Irrespective of whether synchronous state is chaotic or
periodic, strong gap junctions induce synchronization of neurons.

In summary, we have studied synchronous state of a large ensemble of
modified Hodgkin-Huxley (MHH) neurons assuming gap junctions among
neurons.  For the general mean-field dynamics in Eq.~(\ref{eq:general}),
we have evaluated $N\times n$-dimensional deviation to define tangential
Lyapunov exponents~${\{ {\lambda^{\mbox{\tiny $\parallel$}}_{l}}\}}$ and transversal Lyapunov exponents~${\{ {\lambda^{\mbox{\tiny $\perp$}}_{l}}\}}$.
In Fig.~\ref{fig:result}{(b)}, we have investigated characteristic of
synchronous state of MHH neurons by the largest tangential Lyapunov
exponent~${\lambda^{\mbox{\tiny $\parallel$}}_{1}}$. In Fig.~\ref{fig:result}{(c)}, we have elucidated
stability of this synchronization by the largest transversal Lyapunov
exponent~${\lambda^{\mbox{\tiny $\perp$}}_{1}}$.  With weak gap junctions $g=0.02$~{$\mbox{mS/cm}^2$}, stability of
synchronization of MHH neurons shows rather complicated change with
temperature as shown in Fig.~\ref{fig:result}{(c)}.  However, in
Fig.~\ref{fig:largeg} and so on, we have found that with strong gap junctions,
synchronous state is stable over the wide range of temperature.
  The
strong gap junctions induce synchronization both in periodic and chaotic
neurons, and that implies a pivotal role of gap
junctions in synchronization of the large number of neurons

It should be emphasized that the computational cost for Lyapunov
exponents of four-dimensional MHH equations is not much higher than that
of the three-dimensional Lorenz equations.  Even when we assume dozens
of ion channels in neuron dynamics, we would be able to calculate both
Lyapunov exponents within the acceptable computation time.  When
synchronous state is periodic and interactions are infinitesimal ($g\ll
1$), one can use the phase reduction method.  In the present system of
MHH neurons, however, ${\lambda^{\mbox{\tiny $\perp$}}_{1}}$ in Fig.~\ref{fig:largeg} fluctuates
dramatically as $g$ increases, even if synchronous state is
periodic. Since the synchronization property with finite $g$ is quite
different from that with infinitesimal $g$, calculation of transversal
Lyapunov exponents is indispensable in investigating the present system.  In
the real nervous system many types of neurons are interconnected with
chemical synapses.  Some authors model the dynamics of chemical synapse
in the manner as $I_i=(g/N)\sum_j (v_{rev}-v_i)s_j$, where constant
$v_{rev}$ denotes reversal potential and $s_j$ obeys the dynamics
$ds_j/dt=\alpha F(v_j)(1-s_j)-\beta s_j$ with the sigmoidal function
$F(v)=1/(1+\exp[-(v-\theta_{syn})/2])$\cite{perkel,wang,wang2}.  In this
case, synchronous state~${{\bf x}}^\ast$ depends on $g$ since
${{\bf G}}({{\bf x}}^\ast,{{\bf x}}^\ast)$ does not vanish.  Moreover, Jacobi matrices of
${{\bf G}}({{\bf x}}_i,{{\bf x}}_j)$ are not constant but depend on ${{\bf x}}_i$ and ${{\bf x}}_j$.
Our analysis is applicable also to such complicated neural networks
since their dynamics are written in the form of Eq.~(\ref{eq:general}).
Although pulse-coupled neural networks based on threshold-crossing spike
timing cannot be written in the form of Eq.~(\ref{eq:general}), we can
employ the similar analysis by carrying out the decomposition of linear
stability discussed in our previous study\cite{myosioka7}.  In that
study, we have evaluated two types of Floquet matrices to show that
periodic synchronous state of integrated-and-fire (IF) neurons are
stable with only inhibitory chemical synapses.  Interestingly, in the
real nervous system, interneurons are found to be connected with
inhibitory chemical synapses and gap junctions\cite{whittington2}.  It
turns out that networks of interneurons take the extremely ideal
structure to induce synchronization of an ensemble of neurons.  The
present approach of stability analysis is applicable to a wide class of
stability problems in neural networks.  Retrieval state in associative
memory neural networks of spiking neurons can be investigated by the
similar stability analysis\cite{myosioka6}.  More complicated neural
networks including pyramidal neurons and interneurons\cite{kopell} would
also be analyzed by the present approach.

\begin{thebibliography}{99}
\bibitem{buzsaki3} G. Buzs{\'a}ki, Z. Horv{\'a}th, R. Urioste, J. Hetke, and K.~Wise, Science, 256, 1025 (1992).
\bibitem{wang} X.J~Wang and J.~Rinzel, J. Neurosci., 16, 6402 (1996).
\bibitem{golomb} D.~Golomb and J.~Rinzel, Phys. Rev. E, 48, 4810 (1993).
\bibitem{whittington1} M.A.~Whittington, R.D.~Traub, and J.G.R.~Jeffreys, Nature, 373, 612 (1995).
\bibitem{wang2} X.J.~Wang and G.~Buz{\'a}ki, J. Neurosci., 16, 6402 (1996).
\bibitem{chow}  C.C.~Chow, J.A.~White, J.~Ritt, and N.~Kopell, J. Comput. Neurosci., 5, 407 (1998).
\bibitem{tiesinga} P.H.E~Tiesinga and J.V.~Jose, Network 11, 1 (2000).
\bibitem{myosioka7} M.~Yoshioka, Phys. Rev. E, in press.
\bibitem{ermentrout} G.~Ermentrout and N.~Kopell, J. Math. Biol., 29, 195, (1991).
\bibitem{hansel} D.~Hansel, G.~Mato, and C.~Meunier, Neural Comput., 7, 307 (1995).
\bibitem{kuramoto} Y.~Kuramoto, Chemical oscillations, waves, and turbulence (Springer-Verlag 1984).
\bibitem{fujisaka} H.~Fujisaka and T.~Yamada, Prog. Theor. Phys. 69, 32 (1983).
\bibitem{kaneko} K.~Kaneko, Physica D, 77, 456 (1994).
\bibitem{maistrenko} Y.~Maistrenko, T.~Kapitaniak, and P.~Szuminski, Phys. Rev. E, 54, 3285 (1996).
\bibitem{pikovsky} A.~Pikovsky, O.~Popovych, and Yu.~Maistrenko, Phys. Rev. Lett., 87, 044102 (2001).
\bibitem{braun} H.A.~Braun, M.T.~Huber, M.~Dewald, K.~Sch{\" a}fer, and K.~Voigt, Int. J. Bifurcation Chaos
	Appl. Sci. Eng. 8, 881 (1998).
\bibitem{feudel} U.~Feudel, A.~Neiman, X.~Pei, W.~Wojtenek, H.~Braun,
	M.~Huber, and F.~Moss, Chaos, 10, 231 (2000). According to this 
	paper we assume the four-dimensional MMH equations of the form
	$\dot{v}=(-I_l-I_d-I_r-I_{sd}-I_{sr})/c,\ 
	I_l=g_l(v-v_l),\
	I_d=\rho g_d\alpha_{d\infty}(v-v_d),\
	\alpha_{d\infty}=(1+\exp[-s_d(v-v_{0d})])^{-1},\
	I_r=\rho g_r\alpha_{r}(v-v_r),\
	\dot{w}_1=\dot{\alpha}_{r}=\phi(\alpha_{r\infty}-\alpha_{r})/\tau_r,\
	\alpha_{r\infty}=(1+\exp[-s_r(v-v_{0r})])^{-1},\
	I_{sd}=\rho g_{sd}\alpha_{sd}(v-v_{sd}),\
	\dot{w}_2=\dot{\alpha}_{sd}=\phi(\alpha_{sd\infty}-\alpha_{sd})/\tau_{sd},\
	\alpha_{sd\infty}=(1+\exp[-s_{sd}(v-v_{0sd})])^{-1},\
	I_{sr}=\rho g_{sr}\alpha_{sr}(v-v_{sr}),\
	\dot{w}_3=\dot{\alpha}_{sr}=\phi(-\eta{}I_{sr}-\theta\alpha_{sr})/\tau_{sr}$
	with temperature-dependent scaling factors
	$\rho=A_1^{(T-T_0)/10}$ and $\phi=A_2^{(T-T_0)/10}$.
	The parameters values are
	$c=1,\
	v_l=-60,\ v_d=v_{sd}=50,\ v_r=v_{sr}=-90,\
	g_l=0.1,\ g_d=1.5,\ g_r=2.0,\ g_{sd}=0.25,\ g_{sr}=0.4,\
	\tau_r=2.0,\ \tau_{sd}=10,\ \tau_{sr}=20,\
	v_{0d}=-25,\ v_{0r}=-25,\ v_{0sd}=-40,\
	s_d=0.25,\ s_r=0.25,\ s_{sd}=0.09,\
	\eta=0.012,\ \theta=0.17,\
	A_1=1.3,\ A_2=3.0,\ T_0=25$.
\bibitem{shimada} I.~Shimada and T.~Nagashima, Prog. Theor. Phys. 61, 1605 (1979).
\bibitem{milnor} J. Milnor, Commun. Math. Phys. 99, 117 (1985).
\bibitem{whittington2} M.~Whittington and R.D.~Traub, Trends Neurosci. 26, 676 (2003).
\bibitem{myosioka6} M.~Yoshioka, Phys. Rev. E, 66, 061913 (2002).
\bibitem{perkel} D.H.~Perkel, B.~Mulloney, R.W.~Budelli, Neurosci. 6, 823 (1981).\bibitem{kopell} N.~Kopell, G.B.~Ermentrout, M.A.~Whittington, and R.D.~Traub, Proc. Natl. Acad. Sci. USA, 97, 1867 (2000).
\end{thebibliography}
\end{document}